\documentclass[11pt]{article}
\usepackage{amsmath} 
\usepackage{amsthm}
\usepackage{fullpage}
\usepackage{setspace}
\usepackage{maple2e}
\newtheorem{defin}{Definition}
\newtheorem{lemma}{Lemma}
\newtheorem{theorem}{Theorem}
\DeclareMathOperator{\feas}{feas}
\DeclareMathOperator{\LP}{LP}

\DefineParaStyle{Maple Output}
\DefineParaStyle{Maple Plot}
\DefineCharStyle{2D Math}
\DefineCharStyle{2D Output}

\begin{document}

\title{Sufficiently Fat Polyhedra are Not $2$-Castable\thanks{Research
    funded in part by NSERC Canada}}

\author{David  Bremner\thanks{University of New Brunswick.
      \texttt{bremner@unb.ca}} and Alexander Golynski\thanks{University
        of New Brunswick}}

 \maketitle

\begin{abstract}
  In this note we consider the problem of manufacturing a convex
  polyhedral object via casting.  We consider a generalization of the
  sand casting process where the object is manufactured by gluing
  together two identical faces of parts cast with a single piece mold.
  In this model we show that the class of convex polyhedra which can
  be enclosed between two concentric spheres with the ratio of their
  radii less than $1.07$ cannot be manufactured using only two cast
  parts.
\end{abstract}

\section{Introduction}
Casting is a common manufacturing process where some molten substance
is poured or injected into a cavity (called a \emph{mold}), and then
allowed to solidify.  In many applications (see
e.g.~\cite{e-cit-88,i-icmf-87}) it is desirable to remove the cast
object from the mold without destroying the mold (or, obviously the
recently manufactured object).  In general this requires the mold be
partitioned into several parts, which are then translated away from
the cast object.  In the simplest case (prevalent in sand casting), a
mold for polyhedron $P$ is partitioned into two parts using a plane.
If a successful partition (i.e.\ both parts can be removed by
translations without collisions), we say that $P$ is $2$-castable.

Guided by intuition about smooth objects, one might suspect that all
convex polyhedra are $2$-castable. This turns out not to be the case.
Bose, Bremner and van Kreveld~\cite{bbk-dcsp-97} gave an example of a
12 vertex convex polyhedron that is not $2$-castable.  Unfortunately
the proof of non-castability relies on a computer based exhaustive
search. Majhi, Gupta and Janardan~\cite{mgj-cfufp-99} gave a simpler
example with only 6 vertices; here the proof of non-castability is
left as an exercise.  In neither case can one draw any general
conclusions (beyond the tautological) about what sort of convex
polyhedra are $2$-castable.

In the present note we provide a general class of convex polyhedra
that are not $2$-castable.  In particular, we establish that if the
polyhedron has vertices and facets in general position and is a
sufficiently close approximation of a sphere, then it is not
$2$-castable.

\section{Background}

We will actually consider a slightly more general definition of
$2$-castability.  We start with the definition of a castable
polyhedron.  Consider a 3-dimensional half-space $H$.  Let $P$ be a
polyhedron that lies in $H$, while one of its faces $F$ lies on the
boundary of $H$.  The set $M=H \setminus P$ is called a {\em mold} for
$P$.  We say that $P$ is {\em castable with respect to a face} $F$ (or
equivalently {\em with respect to a mold} $M$), if we can pull $P$ out
of $M$ by moving it along some vector $d$ without collisions (e.g.
interior intersections).  If there exists such a face of $P$, then $P$
is called {\em castable}.  A polyhedron $P$ that can can be divided
into $k$ castable parts is called {\em $k$-castable}.  Note that for
$k=2$ it is often required by manufacturing processes that the two
halves are castable with respect to the same mutual facet.  This
constraint is relaxed here, however we do require that the halves are
separated by a plane (i.e. have a mutual facet).

\begin{defin}
\label{fat_def}
A convex polyhedron $P$ is called {\em $(R_i,R_o)$-fat }
if there are two concentric spheres 
$D_i$ and $D_o$ of radii $R_i$ and $R_o$,
such that $D_i \subset P \subset D_o$.
\end{defin}

\begin{defin}
A convex polyhedron is said to be in {\em general position}
if none of its 4 points are coplanar and none of its 3 faces are
parallel to a line.
\end{defin}

Let $P$ be a general position $(R_i,R_o)$-fat polyhedron.
By scaling $P$ we can assume $R_i=1$ and $R_o=R>1$. For the rest of
the paper we use \emph{fat} to mean $(1,R)$-fat.
Let $O$ denote the center of concentric spheres from
Definition~\ref{fat_def}. The following lemma gives bounds
for various elements of a fat polyhedron.

\begin{lemma}
\label{ineq}
Let $P$ be a fat polyhedron.
Every edge of $P$ has length at most
$l^\ast=2\sqrt{R^2-1}$,
every face has area at most 
$S^\ast=\pi(R^2-1)$ and
its volume is bounded
$\frac{4}{3}\pi<V(P)<\frac{4}{3}\pi R^3$.
\end{lemma}
\begin{proof}
Let $AB$ be an edge of $P$ and $O'$ be the projection of $O$ to $AB$.
Since $|OA| \leq R$ and $|OO'| \geq 1$, 
we have $|AO'| \leq \sqrt{R^2-1}$, 
so $|AB| \leq |AO'|+|O'B|< 2\sqrt{R^2-1}$.

Every face $F$ of $P$ defines a slice $C$ of the outer sphere $D_o$,
indeed $F$ is contained within the disk $C$. 
By the previous consideration, the radius of $C$ is at most $l^\ast/2$, 
therefore we have the bound $S(F) \leq \frac{\pi}{4}(l^\ast)^2$ 
on the area of the face $F$.

Since $D_i \subset P \subset D_o$, we have the bounds on the volume of
$P$ given by the lemma.
\end{proof}

The following observation is simple but important for the rest of the paper:
$l^\ast$ can be made arbitrary small by an
appropriate choice of $R$, that is $l^\ast \to 0$ as $R \to 1$.
The following lemma gives an upper bound to the volume of a castable
polyhedron.

\begin{lemma}
\label{cast}
Suppose that $P$ is castable through a face $F$ of area $S$.
Let $H$ be the plane containing $F$ and
$h$ be the maximum distance from a point $P$ to $H$.
Then $V(P) \leq Sh$.
\end{lemma}
\begin{proof}
Let $v$ be the inner normal vector to $F$ and
$F(t)$ be the area of $P \cap H+tv$, for $t \geq 0$.
Since $P$ is castable through $F$, the area $F(t)$ cannot be less then
$S$. Thus $V(P)=\int_0^h{F(t)} \leq Sh$.
\end{proof}
Let us call $h$ the {\em thickness} of $P$ with respect to $F$. 
Note that thickness is bounded by the 
diameter of $P$, thus it cannot exceed $2R$.

\section{The proof of non-castability}

We use the following method to prove that $P$ is not $2$-castable.  We
will consecutively assume that certain polyhedra are castable under
some restrictions and than argue that this implies certain lower
bounds on $R$. Assume that all the possible situations are covered and
let $R>R^\ast$ be the loosest bound on $R$. Then $P$ is not
$2$-castable provided $R < R^\ast$. In the following, let $S(F)$
denote the area of polygon $F$.

\begin{description}
\item[(I)]
First assume that $P$ is {\em 1-castable} through some face $F$.
Using Lemma~\ref{cast} and Lemma~\ref{ineq}, we derive the following
inequality.
\begin{equation}
\pi (R^2-1) 2R \geq S(F) h \geq V(P) > \frac{4}{3}\pi
\label{P}
\end{equation}
This inequality implies the  bound $R>1.240011810$.

\item[(II)]
Suppose that $P$ is {\em $2$-castable}.
Let $P$ be sliced by a plane.
Denote the larger part by $P_1$, the smaller by $P_2$ and
their mutual face by $C$, so that $V(P_1) \geq V(P)/2 \geq V(P_2)$.
To simplify the presentation here, without loss of generality assume
that $O=(0,0,0)$, $C$ is horizontal, 
$P_1$ lies above $C$ and $P_2$ below. 
Let the plane containing $C$ be given by the equation $z=z_0$.
Each of $P_1$ and $P_2$ has to be $1$-castable.

\item[(IIa)]
Assume that $P_1$ is castable through a face $F \neq C$.
As before, we have 
\begin{equation}
\pi (R^2-1) 2R \geq S(F) h \geq V(P_1) 
\geq \frac{V(P)}{2} > \frac{2}{3}\pi
\label{P1}
\end{equation}
Note, that the bound in \eqref{P1} is looser then in \eqref{P}.
The numerical solution gives $R>1.137158043$.

\item[(IIb)]
Assume that $P_1$ is castable through $C$, and
$P_2$ is castable through a face $F \neq C$. 

First consider the case, where $z_0 \geq 0$.
Using Lemma~\ref{cast} for $P_2$ we derive 
\begin{equation}
\label{pos_z}
\pi (R^2-1)2R > S(F) h > V(D_i)/2 = 2\pi/3
\end{equation}
since $P_2$ contain the lower part of the inner sphere.c
Solving \eqref{pos_z} numerically we obtain $R>1.137158043$.

Now consider the case of $z_0 < 0$.
Since $P_1$ contain the disk $D_i \cap \{z=0\}$, we require that the
diameter of the slice $C$ is at least $1$, and hence
\begin{equation}
\label{neg_z1}
\sqrt{R^2-z_0^2}>1
\end{equation}
Using Lemma~\ref{cast} for $P_2$ gives
\begin{equation}
\label{neg_z2}
\pi (R^2-1) 2\sqrt{R^2-z_0^2} \geq S(F) h \geq V(P_2) > 
\pi \int_{-1}^{z_0}{1-t^2 dt}= \pi (2/3+z_0-z_0^3)
\end{equation}
since the diameter of $P_2$ is at most $2\sqrt{R^2-z_0^2}$.
The numerical solution of the system of \eqref{neg_z1} and
\eqref{neg_z2} gives $R>1.07218989$.
\end{description}
We summarize the arguement so far with the following lemma.
\begin{lemma}
If $P$ is $2$-castable then one of the following conditions is true
\begin{itemize}
\item both $P_1$ and $P_2$ are castable through the face $C$.
\item $R>1.07218989$
\end{itemize}
\end{lemma}

Assume that both $P_1$ and $P_2$ are castable
through $C$.
For each edge $e$ of $C$ consider its incident faces $F_1$ in $P_1$ and
$F_2$ in $P_2$ other than $C$.
We \emph{mark} $e$, $F_1$ and $F_2$ if these faces constitute a face of $P$.

\begin{lemma}
There are at most $2$ unmarked edges.
\end{lemma}
\begin{proof}
Each unmarked edge corresponds to an edge of $P$ that lies in $C$.
If we have more than two such edges, then there are at least $4$
vertices of $P$ that lie in $C$, which contradicts the general position 
assumption.
\end{proof}

Consider the set of feasible casting directions
$d=(d_x,d_y,d_z)$ for a polytope $P$.
Without loss of generality assume that 
$d_z=-1$ for $P_1$ and $d_z=1$ for $P_2$.

It is known \cite{compgeom} 
that each face $F$ of $P_1$ ($P_2$) implies a linear constraint on $d$,
namely $(\mu,d) \leq 0$, where $\mu$ is the outward normal to $F$ with
respect to $P_1$ ($P_2$). 
We restrict ourselves to the faces of $P_1$ ($P_2$)
which are incident with marked edges of $C$.
Let $\LP_1$ ($\LP_2$) be the corresponding 2-dimensional linear programs.
Castability of $P_1$ ($P_2$) implies feasibility of $\LP_1$ ($\LP_2$).
Note that if a face $F_1$ of $P_1$ contributes to $\LP_1$ 
then the incident face $F_2$ of $P_2$ contributes to $\LP_2$. 
The corresponding inequalities are:
\begin{eqnarray}
l_1(d) & = &\mu_x d_x+\mu_y d_y-\mu_z \leq 0\\
l_2(d) & = &\mu_x d_x+\mu_y d_y+\mu_z \leq 0
\end{eqnarray}

Let us define $\feas(\LP_i)$ to be the feasible region of a program
$\LP_i$, i.e.\ the intersection of the constraints 
$l_i(d) \leq 0$ for each $\mu$.
We study these programs in more detail. 
First note that feasible regions of these programs have to have inner
points.
For otherwise some there are either:
\begin{itemize}
\item
Two constraints of the form 
$(\mu,d) \leq 0$ and $(-\mu,d) \leq 0$ (their bounding lines
coincide). 
Then the corresponding faces are parallel.
\item
Three constraints, such that their bounding lines on the plane 
$d_z=\pm 1$ intersect in a point.
 This means that the corresponding
faces are parallel to a line.
\end{itemize}

Let $d$ and $e$ be inner points in $\feas(\LP_1)$ and $\feas(\LP_2)$ 
respectively.
Then for every $l_1 \in \LP_1$, we have $l_1(d) < 0$.
So $l_2(-d)=-l_1(d) > 0$ and $-d$ satisfies no constraints of
$\LP_2$.
Consider the ray $r=(-d)+\lambda(e-(-d))$ starting at $-d$ towards $e$. 
The segment between $-d$ and $e$ of this ray has to 
intersect every bounding line in $\LP_2$, 
since one of its endpoints satisfies all the
constraints while the other satisfies none of them. 
Therefore the remainder of $r$ (beyond $e$, $\lambda \geq 1$) 
can not intersect any any of the bounding lines of $\LP_2$.
We conclude that $\feas(\LP_2)$ is unbounded with respect to $r$. 
Similarly we can prove, that $\feas(\LP_1)$ is also unbounded.
Suppose that $\feas(\LP_1)$ is unbounded along a ray $r'=p+\lambda v$,
$\lambda \geq 0$. 

\begin{figure}[ht]
\label{geom}
\begin{center}
\epsfig{file=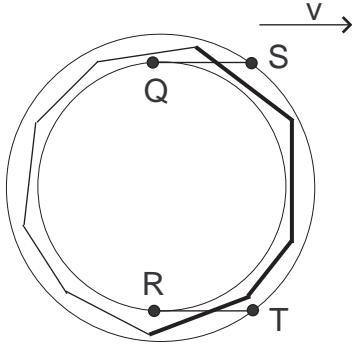,height=5cm}\\
\end{center}
\caption{The face $C$, bold edges are unmarked and form the chain}
\end{figure}
Consider the plane containing the convex polygon $C$ illustrated in 
Figure~\ref{geom}.
Suppose there exists a marked edge $e$ with the outward normal $n$,
such that $(n,v) > 0$. 
Let $\mu$ be the outward normal of the corresponding face of $P$, 
whose projection to $\{z=0\}$ is $n$. 
Then 
$$\mu_x(p_x+\lambda v_x)+\mu_y(p_y+\lambda v_y) + \mu_z = 
\lambda(n,v) +\mu_x p_x + \mu_y p_y + \mu_z \leq 0$$
for every $\lambda \geq 0$, which is a contradiction.
So all marked edges have outward normals $n$ such that 
$(n,v) \leq 0$. 
Consider the edges of $C$ with the outward normal $n$, 
such that $(n,v)>0$. All such edges are unmarked and 
form a chain since $C$ is convex.
Define the segments $QS$ and $RT$ to be the segments parallel to the
vector $v$ and touching the interior circle at the points $Q$ and $R$ 
and the points $S$ and $T$ lie on the exterior circle (see Figure~\ref{geom}).
The first and last edge of $C$, that intersect the interior of  
$QRTS$ have to belong to this chain. 
So the chain connects $QS$ and $RT$, 
hence its length is at least $2$.
But we know that there are at most two unmarked edges, 
thus we one of them has to be longer then $1$.
This means that $l^\ast = 2\sqrt{R^2-1}>1$, an thus 
$R>\sqrt{5/4}>1.118033989$.
Bringing all of the bounds on $R$ together, we conclude with
\begin{theorem}
  Let $P$ be a $(R_i,R_o)$-fat polyhedron in general position
  (i.e.\ no four vertices of which lie on a plane and no three faces
  are parallel to a line). Then $P$ is not $2$-castable if
  $R_o/R_i<1.07218989$.
\end{theorem}

\bibliographystyle{plain}
\bibliography{casting}

\begin{thebibliography}{1}

\bibitem{bbk-dcsp-97}
Prosenjit~K. Bose, David Bremner, and Marc van Kreveld.
\newblock Determining the castability of simple polyhedra.
\newblock {\em Algorithmica}, 19(1--2):84--113, September 1997.

\bibitem{e-cit-88}
R.~Elliott.
\newblock {\em Cast Iron Technology}.
\newblock Butterworths, London, 1988.

\bibitem{i-icmf-87}
A.I. Isayev, editor.
\newblock {\em Injection and Compression Molding Fundamentals.}
\newblock Marcel Dekker, Inc., New York, 1987.

\bibitem{compgeom}
M.~Overmars M.~de Berg, M. van~Kreveld and O.~Schwarzkopf.
\newblock {\em Computational Geometry: Algorithms and Applications}.
\newblock Springer-Verlag, 1997.

\bibitem{mgj-cfufp-99}
J.~Majhi, P.~Gupta, and R.~Janardan.
\newblock Computing a flattest, undercut-free parting line for a convex
  polyhedron, with application to mold design.
\newblock {\em Comput. Geom. Theory Appl.}, 13:229--252, 1999.

\end{thebibliography}

\end{document}